\documentclass [11pt, letterpaper]{article}
\pdfoutput=1 
\usepackage{fullpage}
\usepackage{amsmath}
\usepackage{amssymb}
\usepackage{graphicx}
\usepackage{mathrsfs}
\usepackage{wrapfig}
\usepackage{boxedminipage}
\usepackage{epsfig}
\usepackage{setspace}
\usepackage{subfigure}
\usepackage{float}
\usepackage{hyperref}
\usepackage{enumerate}
\usepackage{cite}
\usepackage[font=footnotesize,labelfont=rm]{caption}

\begin{document}

\begin{flushright}
{\tt arXiv:1404.5982}
\end{flushright}

{\flushleft\vskip-1.35cm\vbox{\includegraphics[width=1.25in]{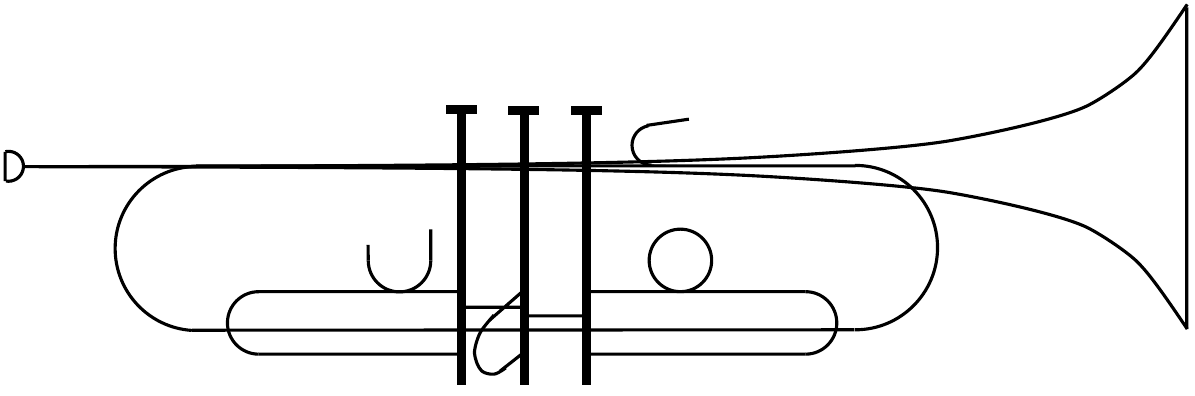}}}

\bigskip
\bigskip
\bigskip
\bigskip

\bigskip
\bigskip
\bigskip
\bigskip

\begin{center} 

{\Large\bf  Holographic  Heat Engines}

\end{center}

\bigskip \bigskip \bigskip \bigskip

\centerline{\bf Clifford V. Johnson}

\bigskip
\bigskip
\bigskip

  \centerline{\it Department of Physics and Astronomy }
\centerline{\it University of
Southern California}
\centerline{\it Los Angeles, CA 90089-0484, U.S.A.}

\bigskip

\centerline{\small \tt johnson1,  [at] usc.edu}

\bigskip
\bigskip


\begin{abstract} 
\noindent It is shown that in theories of gravity where the cosmological constant is considered a thermodynamic variable, it is natural to use  black holes  as heat  engines. Two examples are presented in detail using AdS charged black holes as the working substance. We notice that for static black holes, the maximally efficient traditional Carnot engine is  also a Stirling engine. The case of negative cosmological constant supplies  a natural realization of these engines  in terms of  the field theory description of the fluids to which they are holographically dual.  We first propose a precise picture of how the traditional thermodynamic dictionary of holography is extended when the cosmological constant is dynamical and then conjecture that the engine cycles can be  performed by using renormalization group flow. We speculate about the  existence of  a natural dual field theory counterpart to the gravitational thermodynamic  volume. 
\end{abstract}
\newpage \baselineskip=18pt \setcounter{footnote}{0}

\section{Extended Black Hole Thermodynamics}
%

Recently, the  classic subject of black hole thermodynamics\cite{Bekenstein:1973ur,Bekenstein:1974ax,Hawking:1974sw,Hawking:1976de}, which relates the mass $M$, surface gravity $\kappa$, and area $A$ of a black hole to the energy $U$, temperature $T$, and entropy $S$, according to:
\begin{equation}
M=U\ ,\quad T=\frac{\kappa}{2\pi}\ , \quad S=\frac{A}{4}\ ,
\end{equation}
has been extended\footnote{For a selection of references, see refs.\cite{Caldarelli:1999xj,Wang:2006eb,Sekiwa:2006qj,LarranagaRubio:2007ut,Kastor:2009wy,Dolan:2010ha,Cvetic:2010jb,Dolan:2011jm,Dolan:2011xt}, including the reviews in refs.\cite{Dolan:2012jh,Altamirano:2014tva}. See also the early work in refs.\cite{Henneaux:1984ji,Teitelboim:1985dp,Henneaux:1989zc}.} to include black hole counterparts for the pressure $p$ and volume $V$. The  cosmological constant of the spacetime in question supplies the pressure {\it via} $p=-\Lambda/8\pi$, while the thermodynamic volume $V$ is associated with the volume occupied by the black hole itself\footnote{The term ``associated'' is used since there is a subtlety to be discussed later.}. (Here we are using geometrical units where $G,c,\hbar,k_{\rm B}$ have been set to unity. We may restore them using dimensional analysis when required later.)  The formalism works in multiple dimensions, and our remarks will apply to those situations too, although for clarity we will  mostly write four--dimensional  formulae.The black holes may have other parameters such as gauge charges $q_i$ and angular momenta $J_i$, and these, with their conjugates the potentials  $\Phi_i$ and angular velocities $\Omega_j$, enter additively into the first law in the usual manner. 

In the presence of a variable pressure $p$ (now identified with the cosmological constant), ref.\cite{Kastor:2009wy} proposed that the extension shifts the identification of the mass $M$  from being the energy~$U$ to  being the {\it enthalpy}, to wit: $M=H\equiv U+pV$, so the First Law now becomes:
\begin{equation}
dM=TdS+Vdp+ \Phi dq+ \Omega dJ\ ,
\end{equation}
in four dimensions with an electric charge and rotation. When $p$ is removed from the list of variables, we return to the usual situation.

In the case of static black holes, the thermodynamic volume $V$ is simply the ``geometric'' volume constructed by naive use of the radius of the black hole horizon to form the associated volume\footnote{Such a definition agrees with the definition of the volume of a static black hole proposed in ref.\cite{Parikh:2005qs}.}. For example, in four dimensions,  for a Schwarzschild or Reissner--Nordstr\"om  black hole with horizon radius  $r_h$, we have 
\begin{equation}
V=\frac{4}{3}\pi r_h^3\ .
\end{equation}
Enthalpy is very natural here: The cosmological constant is a spacetime energy density of  $-p=\Lambda/8\pi$ per unit volume. Forming a black hole of volume $V$ requires cutting out a region of spacetime of that volume, at cost $pV$,  and this energy of formation is naturally captured by the enthalpy. It is important to note that the entropy $S$ is already related to  the horizon radius through its relation to area {\it via} the Bekenstein area law. So in this case of static black holes, the thermodynamic volume~$V$ and the entropy $S$ are simply related to each other. This is key to the simplicity of one of the  results concerning thermodynamic cycles  presented below.  The lack of independence of $S$ and $V$ would be a concern if   studying problems  that use  the internal energy $U(S,V)$  as the central thermodynamic potential, but we are  in a situation where it is the enthalpy $H(S,p)$ that is  natural. Pressure and entropy are the key variables here, and they are independent  for the holes in question.

Note that for rotating black holes, the thermodynamic volume $V$ and the entropy $S$ are independent (the situation is resolved by non--zero angular momentum $J$), and there are no special subtleties involving $U$ as a result. In fact, the thermodynamic volume is no longer the naive geometric volume occupied by the black hole in this case. Refs.\cite{Caldarelli:1999xj,Cvetic:2010jb,Dolan:2011xt}  expand upon these issues.

\section{Thermodynamic Cycles and Heat Engines}
\label{sec:engines}
With pressure and volume in play alongside temperature and entropy,  the possibility of extracting mechanical useful work from heat energy naturally springs to mind. (We may also consider heat pumps or refrigerators, where instead  work  is done to transfer heat from a cold reservoir to a hot one. The flows  in the cycles to be discussed  may simply be reversed to cover those cases.)  
 
\begin{wrapfigure}{L}{0.3\textwidth}
{\centering
\includegraphics[width=1.8in]{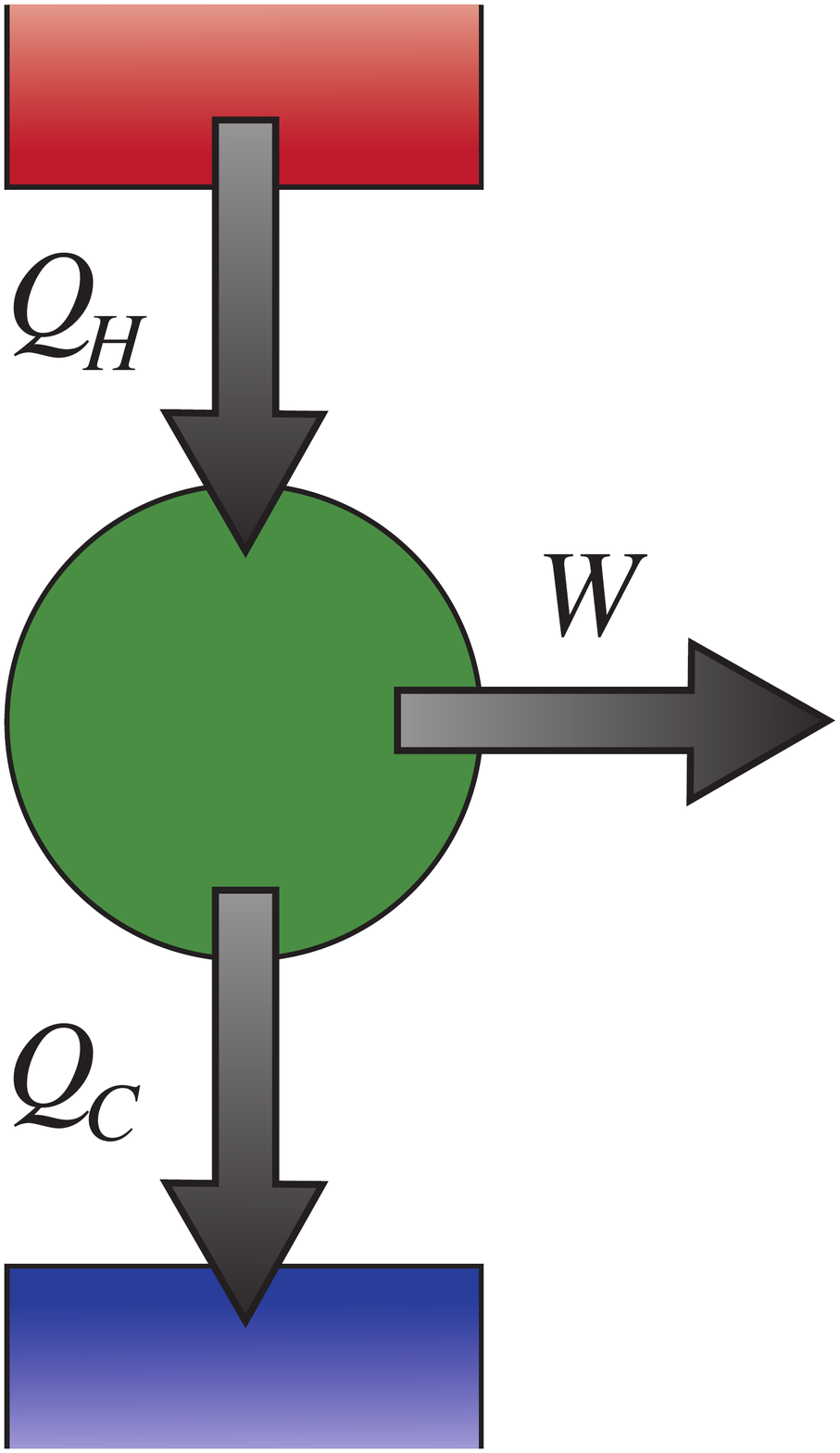} 
   \caption{\footnotesize   The  heat engine flows.}   \label{fig:heatengine}
}
\end{wrapfigure}
\noindent
We  can start with  an equation of state, {\it e.g,} a function  $p(V,T)$, and define an engine as a closed path in the  $p{-}V$ plane, allowing for the input of an amount of heat $Q_H$, and the exhaust of an amount~$Q_C$. The total mechanical work done, by the First Law, is of course $W=Q_H-Q_C$. So  the efficiency of the heat engine is defined to be $\eta= W/Q_H = 1-Q_C/Q_H$. Figure~\ref{fig:heatengine} shows the standard logic of the energy flows for one cycle of the engine.

The precise engine we make depends upon the choice of path in the $p{-}V$ plane, and possibly the equation of state of the black hole in question. Let us make a simple cycle as follows: Some of the classic cycles involve a pair of isotherms at temperatures $T_H$ and~$T_C$,  where $T_H>T_C$, where there is an isothermal expansion while some heat is being absorbed, and an isothermal compression during the expulsion of  some heat. We can connect these in a variety of ways, but two simple choices are natural. We can do isochoric paths to connect the two temperatures, as in the classic Stirling cycle, or we can do adiabatic paths, as in the classic Carnot cycle.  For the latter, all the heat flows of the engine take place during those two isotherms, and (with the usual assumption that we do things slowly enough to be in the quasistatic regime) these are  reversible. The whole heat engine is fully reversible (since the total entropy flow is zero) and so  the engine should have the  Carnot efficiency, which is set simply by the temperature difference: $\eta=1-T_C/T_H$. This the maximum efficiency any heat engine can have when operating between these temperatures. Any higher efficiency would violate the Second Law.

This is therefore the gold standard engine, and so it is interesting to explore how it is precisely realised in black hole thermodynamics since any other black hole  heat  engine that might be made will be measured against this one.  Now, it is comforting to keep in mind  that whatever the equation of state, the above described Carnot path will yield the Carnot efficiency\footnote{This is a general result in thermodynamics following from the Second Law and the vanishing of the net entropy flow. For a recent alternative interesting  ({\it i.e.} not directly appealing to the Second Law) proof, see ref.\cite{recentpaper}.}, but nevertheless it is important and useful to know exactly what the shape of the paths are for a given system.  For a general black hole, working out the explicit equation of state can be a difficult task (it is usually easier to define $p$, $V$, and $T$ in terms of another natural variable such as the horizon radius, or the entropy, which implies the equation of state upon elimination of that intermediate variable), and it is additionally complex (for a sufficiently complicated equation of state) to have a closed form equation for both the isotherms and adiabats. So it is  a daunting task to determine the shapes of the Carnot cycle for the black holes explicitly.  

This is where, for the static holes, the fact that the thermodynamic volume $V$ and the entropy $S$ are not independent is key. It means that adiabats and isochores are the same! Carnot and Stirling coincide. So the efficiency  of our cycle  may be simply computed, and some of the path known explicitly, without knowledge of the detailed equation of state. All that's needed is that the entropy and volume are related.

\begin{wrapfigure}{l}{0.45\textwidth}
{\centering
\includegraphics[width=2.1in]{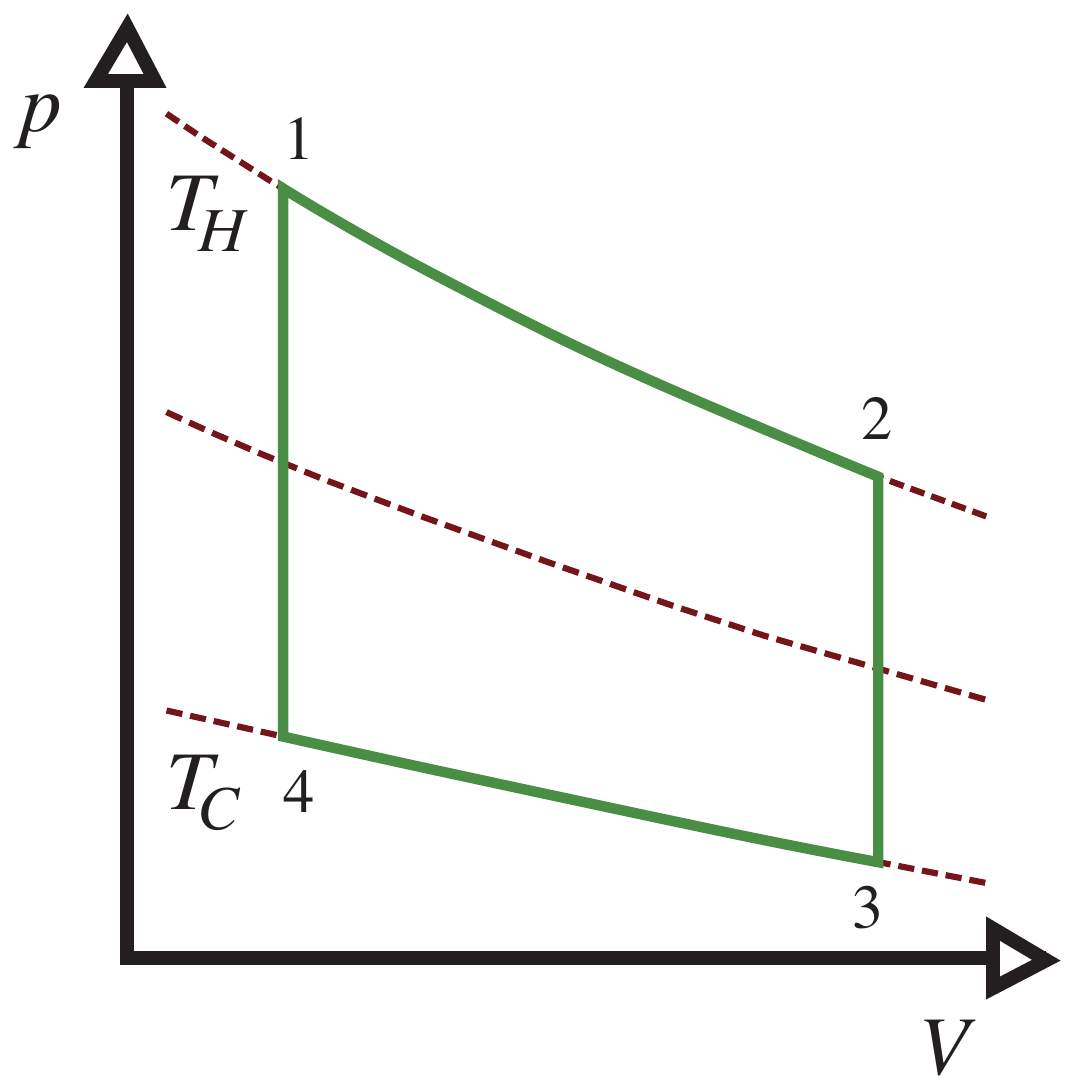} 
  \caption{\footnotesize  Our Carnot engine, which for static black holes is also a Stirling Engine.\newline}  
  \label{fig:cyclesa}
}
\end{wrapfigure}

So along the upper isotherm (subscripts  refer to the labelling in figure~\ref{fig:cyclesa}) we have 
the following heat flow: 
\begin{equation}
Q_H=T_H\Delta S_{1\to2}=T_H\left(\frac{3}{4\pi}\right)^{\frac23}\pi \left(V_2^{\frac23}-V_1^{\frac23}\right)\ ,
\end{equation}
and along the lower:
\begin{equation}
Q_C=T_C\Delta S_{3\to4}=T_C\left(\frac{3}{4\pi}\right)^{\frac23}\pi \left(V_3^{\frac23}-V_4^{\frac23}\right)\ . 
\end{equation}
Since $V_1=V_4$ and $V_2=V_3$ (we moved along isochores),  the efficiency becomes:
\begin{equation}
\eta=1-\frac{T_C}{T_H}\ . 
\label{eq:carnot}
\end{equation}
Happily, for static black holes the equation of state can be made explicit too (as we will show in the example of the next section) and so the full shape of the Carnot  cycle for these cases can be fully characterized.  We can make Carnot engines for non--static black holes too, but now the adiabats will not be isochores, and the full equation of state must be used to determine the shapes of the paths.  Using isochores will give the Stirling engine which will have a lower efficiency than Carnot since there'll be additional (non--reversible) heat flows.  

Notice that the  engines can also include  non--trivial phase structure somewhere along the path we chose.  If a phase transition between large and small black holes occurs as the pressure varies along the isotherm, as is well known to take place for such holes\cite{Chamblin:1999tk,Chamblin:1999hg} (see the example below), the Carnot result is robust since all it relies on are the volume differences. It  does not matter whether those differences took place as a result of a discontinuous jump (as in a first order transition) or the milder change of a critical point (as in a second order transition).

\section{An Example: Charged Black Holes in AdS$_4$}
\label{sec:example}
\begin{wrapfigure}{l}{0.6\textwidth}
{\centering
\includegraphics[width=3.8in]{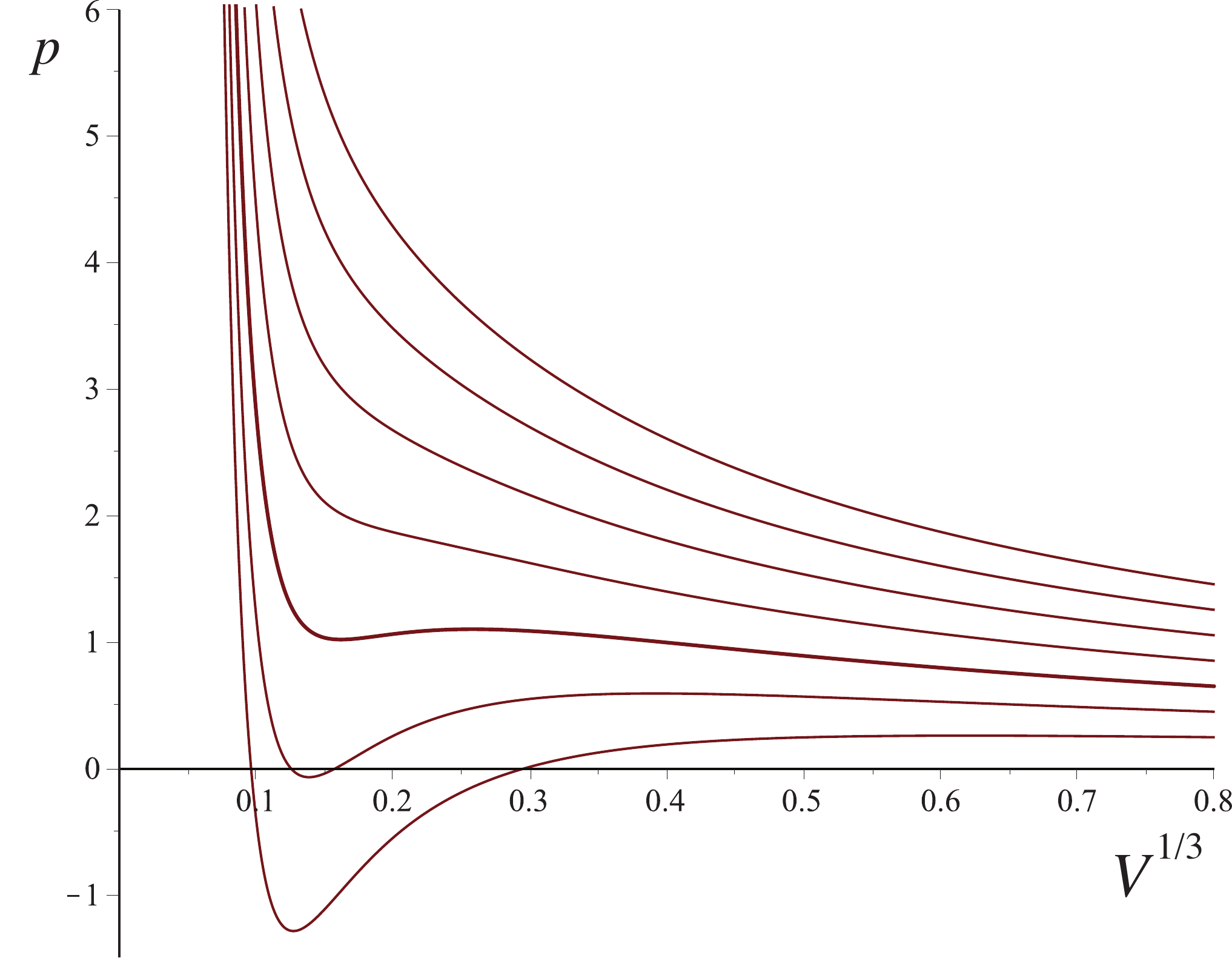} 
   \caption{\footnotesize  Sample (uncorrected) isotherms.  Values chosen were  $q=0.05$ and $T$ from 0.4 to 1.6 in intervals of 0.2. Lower curves are at lower temperatures. The multi--valued parts of the low temperature curves are removed by first order phase transitions, also removing the unphysical negative pressure. See text.}   \label{fig:equationofstate}
}
\end{wrapfigure}

Just for clarity, it is worth exhibiting a concrete example that has all the elements we've discussed, so let us take static black holes in four dimensions with negative cosmological constant. The black hole is a Reissner--Nordstr\"om solution of the Einstein--Maxwell system with bulk action
\begin{equation}
I=-\frac{1}{16\pi }\int \! d^4x \sqrt{-g} \left(R-2\Lambda -F^2\right)\ ,
\label{eq:action}
\end{equation}
where $\Lambda=-3/\ell^2$, the cosmological constant, sets a length scale $\ell$. The black hole has mass and charge  $M$ and ${q}$, with metric\footnote{We've chosen to work with a spacetime which is asymptotic to global AdS in this example. Our general remarks in this paper are not restricted to such situations, and choices for AdS with  flat or hyperbolic slicings are also relevant.}
\begin{equation}
ds^2 = -Y( r)dt^2
+ {dr^2\over Y(r)} + r^2 (d\theta^2+\sin^2\theta d\varphi^2)\ ,
\quad {\rm where}\quad
Y( r) = 1-\frac{2M}{r}+\frac{q^2}{r^2}+\frac{r^2}{\ell^2}\ ,
\end{equation}
and  there is a gauge potential that is chosen to vanish on the horizon located at $r=r_+$, the largest positive real root of $Y(r)$:
$A_t = q(r-r_+)/{rr_+} .$
\noindent
The  requirement of regularity of the Euclidean section fixes the temperature $T$ according to:
\begin{equation}
\frac{1}{T}=4\pi Y^\prime \left.\right|_{r=r_+} = \frac{4\pi \ell^2 r_+^3}{3r_+^4+l^2r_+^2-q^2 \ell^2} \ , 
\label{eq:equationofstate}
\end{equation}
and the entropy is $S=\pi r_+^2$. We can define the pressure $p=3/(8\pi \ell^2)$ and re--arrange the temperature expression above into an equation of state\cite{Chamblin:1999tk,Kubiznak:2012wp} for a given charge $q$:
\begin{equation}
p=\frac{1}{8\pi}\left(\frac{4\pi}{3}\right)^\frac43\left(\frac{3T}{V^\frac13} - \left(\frac{3}{4\pi}\right)^{\frac23}\frac{1}{V^\frac23}+\frac{q^2}{V^\frac43}\right)\ ,
\label{eq:pressure}
\end{equation}
where we substituted $r_+$ for  the thermodynamic volume using  $V=4\pi r_+^3/3$. For our heat engine discussion, since we are interested in the mechanical work, we will fix at a specific value of the  charge and so we will turn off the $\Phi dq$  term, leaving:
\begin{equation}
dH=dM=TdS+Vdp\ .
\end{equation}
The function $H(S,p)$ can be easily computed (for example by converting  the potentials computed in the action computations of ref.\cite{Chamblin:1999tk,Chamblin:1999hg}, or by other methods --- see {\it e.g.,} ref.\cite{Dolan:2011xt}):
\begin{equation}
H(S,p)=\frac{1}{2}\sqrt{\frac{S}{\pi}}\left(1+\frac{\pi q^2}{S}+\frac{8Sp}{3}\right)\ ,
\end{equation}
from which we can recover $V$ and $T$ by partial differentiation. It is easy to check that the consistency conditions for $dH$ to be exact (Maxwell's relations) are satisfied:
\begin{equation}
\left(\frac{\partial T}{\partial p}\right)_{S} = \left(\frac{3V}{4\pi}\right)^\frac13=2\left(\frac{S}{\pi}\right)^\frac12=\left(\frac{\partial V}{\partial S}\right)_p\ .
\end{equation} 
Figure~\ref{fig:equationofstate} shows some sample (uncorrected) isotherms in the  $p{-}V^{\frac13}$  plane. (The structure in the $p{-}V$ plane is of course the same, but much more horizontally stretched.)   
We use the qualifier ``uncorrected'' above for the following reason: As discovered in ref.\cite{Chamblin:1999tk}  (and studied
 with variable pressure  in ref.\cite{Kubiznak:2012wp}), this fixed charge ensemble has  phase transitions due to the multi--valued nature of the equation of state that appears at low enough temperature. There is a first order phase transition between small and large black holes  (as pressure is changed) reminiscent of the Van der Waals liquid--gas system. The resulting jumps between large and small black holes corrects the naive isotherms given by equation~(\ref{eq:pressure}), and removes the negative pressure regions that are evident in figure~\ref{fig:equationofstate}. The resulting line of first order transitions in the $(p,T)$ plane ends in a second order critical point.  Many properties of these transitions have been worked out in refs.\cite{Chamblin:1999tk,Chamblin:1999hg,Kubiznak:2012wp}, and they  won't be a focus here.
 
 An important result that underlies the simple observation that the isochores are adiabats can be derived from first writing the temperature in equation~(\ref{eq:equationofstate})  in terms of $S$ and $p$ as follows:
 \begin{equation}
 T=\frac{1}{4\sqrt{\pi}}\frac{1}{\sqrt{S}}\left(1-\frac{\pi q^2}{S}+8pS\right)\ . 
 \end{equation}
 Then differentiation gives the specific heat:
 \begin{equation}
C= T\frac{\partial S}{\partial T} = \left(1-\frac{2S^\frac12}{\sqrt{\pi}} \frac{\partial p}{\partial T} \right)2S\left(\frac{8pS^2+S-\pi q^2}{8pS^2-S+3\pi q^2}\right)\ ,
 \end{equation}
 which  shows (since $(\partial p/\partial T)_V=\pi^\frac12/2S^\frac12$) that the specific heat at constant volume vanishes $C_V=0$, while $C_p$ is given by setting $\partial p/\partial T=0$ in the expression above\cite{Dolan:2010ha,Kubiznak:2012wp}. The vanishing of $C_V$ is the ``isochore equals adiabat'' result, specific to static black holes, making our Carnot cycles particularly simple to make explicit. We can put a Carnot cycle  on the diagram by picking two isotherms for $T_H$ and~$T_C$, and then dropping two vertical lines between them to close the loop as we did in figure~\ref{fig:cyclesa}.  The loop can  include the jumps in volume as the pressure changes  along an isotherm.
 
 \begin{wrapfigure}{r}{0.43\textwidth}
{\centering
\includegraphics[width=2.3in]{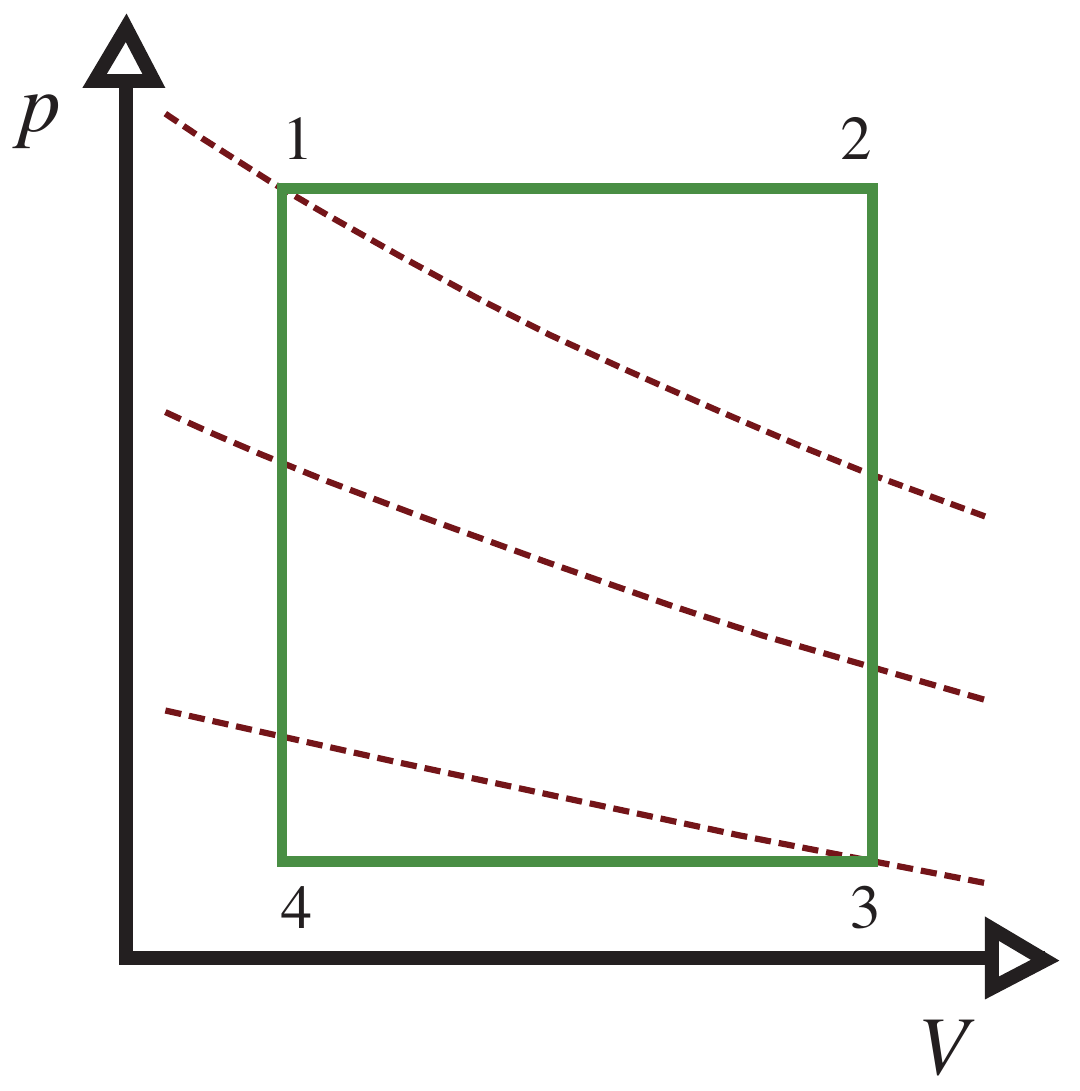} 
   \caption{\footnotesize  Our  other engine.}   \label{fig:cyclesb}
}
\end{wrapfigure}

Actually, an explicit expression for $C_p$ would suggest that we ought to have a new engine that we can analyze simply, involving two isobars and two isochores/adiabats. See figure~\ref{fig:cyclesb}. The work done along the isobars is very easy to compute:
 \begin{equation}
 W= \frac{4}{3\sqrt{\pi}}\left(S_2^\frac32-S_1^\frac32\right)(p_1-p_4)\ ,
 \label{eq:nicework}
 \end{equation}
 where the subscripts refer to the quantities evaluated at the corners labeled (1,2,3,4) and we've written the volume in terms of the entropy to reduce the number of variables in the final expression for the efficiency. The heat flows take place  along the top and bottom. The upper isobar will give the  net inflow of heat,  which is therefore $Q_H$, so we may write:
 \begin{equation}
 Q_H=\int_{T_1}^{T_2} C_p(p_1,T) dT\ ,
\end{equation}
 where the non--trivial entropy dependence of $C_p$ gives a non--trivial $T$ dependence, which makes the integral  messy. In any case, the efficiency is then $\eta=W/Q_H$, where the previous two quantities can be substituted. As a check on our methods we can take a limit where the cycle is at high pressure and temperature. 
 Then our expressions simplify and allow us to perform the integral. We can focus on the large volume branch of solutions and therefore neglect $q$ to leading order, expanding at large $T$ and $p$ to get:
\begin{equation}
S= \frac{\pi}{4}\frac{T^2}{p^2}-\frac{1}{4p}-\frac{1}{16\pi T^2}+\cdots\ ,\quad C_p=\frac{\pi}{2p^2}T^2+\frac{1}{8\pi T^2}+\cdots, 
\end{equation} 
which yields
\begin{eqnarray}
Q_H&=& \frac{\pi}{6p_1^2} (T_2^3-T_1^3)+\frac{1}{8\pi}\left(\frac{1}{T_1}-\frac{1}{T_2}\right)+\cdots\nonumber\\
&=&\frac{4}{3\sqrt{\pi}} p_1\left(S_2^\frac32-S_1^\frac32\right)+\frac{1}{2\sqrt{\pi}}\left(S_2^\frac12-S_1^\frac12\right)+O\left(\frac{1}{p_1^2}\right)\ .
\end{eqnarray}
So dividing  the work in equation~(\ref{eq:nicework}) by this, we have our expression for the efficiency, which we can write as:
\begin{equation}
\eta=\left(1-\frac{p_4}{p_1}\right)\left\{1-\frac38\frac{1}{p_1}\left(\frac{S_2^\frac12-S_1^\frac12}{S_2^\frac32-S_1^\frac32}\right)+O\left(\frac{1}{p_1^2}\right)\right\}\ .
\end{equation} At leading order, since  $p\sim T/V^\frac13+\cdots$ in this limit,   we can see that the efficiency becomes $\eta=1-\frac{T_C}{T_H}\left(\frac{V_2}{V_4}\right)^\frac13$, where we use $T_C=T_4$ and $T_H=T_2$ since those are the lowest and highest temperatures the engine will need to operate at in order to exchange heat with its environment. So we can approach the Carnot efficiency~(\ref{eq:carnot}) at leading order only if we also make the cycle an extremely narrow rectangle (which in the limit would produce no work at all). This makes sense since the  heat exchanges (along the isobars that change the volume to perform the work) are irreversible and comparable in magnitude to the work done, even at high pressure and temperature. Of course, the corrections to the high pressure and temperature limit shown in the expression take us even further away from the ideal.
 
\section{Renormalization Group Engineering}

While our remarks apply to both positive and negative cosmological constant, at least formally, we can imagineer the kind of engine proposed in section~\ref{sec:engines} quite naturally  in the case of negative cosmological constant $\Lambda$, since we have the AdS/CFT holographic correspondence\cite{Maldacena:1997re, Gubser:1998bc, Witten:1998qj, Witten:1998zw}  to help us. The black hole in $D$--dimensional gravity is dual to a non--gravitational field theory of a fluid in $D-1$ dimensions. 

However, first we must solve a puzzle. In the extended thermodynamics we've been discussing, how are we to interpret the pressure and the volume in the dual field theory? Are they the pressure and volume of the field theory? There does not seem to be room for this to work, as can be seen by  studying the stress tensor of the Schwarzschild black hole in AdS\cite{Balasubramanian:1999re}. The stress tensor's  properties are consistent with that of a conformally invariant fluid with density $\rho$ proportional to pressure (see ref.\cite{Johnson:2003gi} for a review). Both are set by the energy (mass $M$ of the black hole, plus the Casimir energy, if we're in global AdS). So that fluid pressure is {\it not} the $p$ of the AdS thermodynamics that is set by the cosmological constant. They simply do not match. 

As it stands, therefore, we have  the standard black hole thermodynamics of the gravity theory, where ($M$, $T$,  $S$)  map (after putting in the value of Newton's constant $G$) to  ($U$, $T$,  $S$) of the dual non--gravitational theory. This is the translation  that is used in standard holographic discussions. On the other hand, we have the extended black hole thermodynamics where $p$ and $V$ are dynamical, and then $M$  is the enthalpy $H=U+pV$ of the gravity theory instead. Should we use this instead for discussing holography?  They agree only when $p$ is not a thermodynamic variable. However, they seem to contradict each other otherwise.  Which system is correct? Is the conclusion that we should never have $p$ dynamical in holographic discussions? This is an issue that does not seem to have been addressed in the literature, and we now propose a resolution.

There is a way that we {\it can} extend holography to include dynamical $p$,  by recognising that  both relations can be correct at the same time.  The mass of the black hole $M$ remains as the  energy $U$ in the dual field theory, but it is {\it also}  the enthalpy $H=U+pV$ in the gravity theory. On the gravity  side, $p$ is dynamical and plays the role of a pressure, while on the non--gravitational  side, although it has meaning, it is not a thermodynamic variable. The same relationships  will be true for the other thermodynamic potentials. The Euclidean path integral $I^{\rm E}/\beta=-\log Z_{\rm grav}/\beta$,  in the usual (fixed $\Lambda$) gravitational thermodynamics  (here, $\beta=1/T$) is to be identified with the   Helmholtz free energy $F=U-TS$ of the dual field theory.  When we allow the cosmological constant ($p$) to vary, the natural quantity it equates to on the gravity side should be the  Gibbs free energy $G=U-TS-pV$.  See table~\ref{tab:potentials} for a summary.
\begin{table}[h]
\begin{center}
  \begin{tabular}{ | l || c | c |}
    \hline
    Gravitational quantity &   $M$& $I^{\rm E}/\beta=-\log Z_{\rm grav}/\beta$ \\ \hline
    Field theory thermodynamic potential &  $U$ & $F= U - TS$ \\ \hline
    Dual gravity  thermodynamic potential & $H=U+pV$ & $G=F+pV=H-TS$ \\ \hline
  \end{tabular}
\end{center}
\caption{\footnotesize Table showing two key quantities computed in the gravity theory and  the thermodynamic potentials  in field theory and gravity that they should correspond to  when the gravity theory has $p$ as a thermodynamic variable (the  ``extended'' thermodynamics).}
\label{tab:potentials}
\end{table}

So on the   field theory side,  what is the meaning of $p$, given that it is not a thermodynamic variable?
The answer remains what it is in the standard AdS/CFT dictionary. In the extended thermodynamics  $p=-\Lambda/8\pi G$ (where here $G$ is Newton's constant and not Gibbs of the previous paragraph) and  $\Lambda=-(D-2)(D-1)/2\ell^2$, in $D$ dimensions. Recall that the value  of the length scale~$\ell$  is set by the Planck length of the underlying uncompactified  theory  ({\it e.g.}, the eleven dimensional Planck length  or ten dimensional string length) and in the simplest examples, a pure number, $N$, related to the number of coincident branes (M--branes or D--branes).  Larger $N$ means larger $\ell$, and as is well known the gauge/gravity correspondence becomes very useful for large $N$, where the curvatures are small. 
\begin{wraptable}{R}{0.28\textwidth}
{\centering
  \begin{tabular}{ | c || c | c | c |}
    \hline
    $D$ & 4 & 5 & 7  \\ \hline
    $\ell$ & $N^{\frac16}$ &   $N^{\frac14}$& $N^{\frac13}$  \\ \hline
    $G$ & $N^{-\frac76}$ &    $N^{-\frac54}$& $N^{-\frac43}$   \\ \hline
    $p$ & $N^{\frac56}$ &   $N^{\frac34}$& $N^{\frac23}$   \\ \hline
  \end{tabular}
}
\caption{\footnotesize Table showing the $N$ dependence of the length scale $\ell$ set by the cosmological constant,  Newton's  constant, and the pressure $p$,  in $D$ dimensions.}
\label{tab:Ndependence}
\end{wraptable}On the field theory side, $N$ is typically the rank of a gauge group of the theory, and as such it also determines the maximum number of available degrees of freedom. (For example, for $U(N)$ it would be $N^2$.)   Table~\ref{tab:Ndependence} gives a summary of the~$N$ dependences in the simplest cases of AdS$_D$ ($D=4,5,7$),  where the $N$ dependence of Newton's constant $G$ in those dimensions (obtained by dimensional reduction) is shown since it is needed to compute the pressure {\it via}: $p=-\Lambda/8\pi G$. Overall, we see that  the pressure $p$, scales with~$N$. So dynamical  $p$ must mean\footnote{Note that although they did not pursue the connection, refs.\cite{Kastor:2009wy,Dolan:2013dga} also mention the link between $p$ and $N$, and wondered as to its significance. We thank B. Dolan and D. Kastor for pointing this out after the first version of this manuscript appeared.} dynamical $N$.

From the perspective of the $D$--dimensional  gravity theory under discussion, $\Lambda$  (and hence the positive pressure $p$) is set by  the value of the potential ${\cal V}(\varphi_i)$ of the scalars $\varphi_i$ of the gravity theory. The pure AdS case is the highly symmetric fixed point  of the gravity theory where $\varphi_i=0$, but there are other fixed points of the potential corresponding to other AdS spaces. They have different values of the potential, and hence different values of $\Lambda$ --- {\it i.e.,} different values for the effective $N$ measuring the available degrees of freedom.  This is the core idea in the holographic renormalization group\cite{Girardello:1998pd,Distler:1998gb} (see ref.\cite{Johnson:2003gi} for a review), and there are many explicit examples, although they are hard to construct (even at zero temperature) in general since the scalar dynamics are highly non--linear. So turning on relevant operators in the field theory (dual to the scalars $\varphi_i$), driving the theory to new IR fixed points, is one way to dynamically change $\Lambda$.

However, in order to explore the full space of available values of negative cosmological constant (and hence of positive $p$) will require more than just turning on relevant operators to trigger flows to IR fixed points. We expect that irrelevant operators will be needed as well (dual to higher mass Kaluza--Klein fields on the gravity side), allowing us to explore the extremes of the Coulomb branch (and the finite temperature deformation thereof) corresponding to moving some number of branes off the collection of $N$ coincident branes whose throat is the AdS$_D$ and moving them off to infinity, or {\it vice--versa}, thereby changing $N$. (This can also be discussed in terms of changing the number of units of flux on the compact directions in the dimensionally reduced picture.) Such operators have been discussed in the literature (mostly at zero temperature) where they correspond to motions on the Coulomb branch that go beyond what is accessible by holographic RG flow with relevant operators\footnote{See {\it e.g.},  refs.\cite{Kraus:1998hv,Klebanov:1999tb,Intriligator:1999ai,Costa:1999sk,Costa:2000gk} for some early discussion at zero temperature, refs.\cite{Skenderis:2006uy,Skenderis:2006di} for later clarification of the connection between gravity and field theory for this sector, and ref.\cite{Evans:2001zn} for some exploration of the finite temperature physics of the kind of irrelevant operator involved. The early work on Coulomb branch solutions accessible by holographic RG flow triggered by low dimension operators are refs.\cite{Freedman:1999gk,Brandhuber:1999jr}. The solutions correspond to smeared distributions of branes in higher dimensions\cite{Kraus:1998hv}. Our application needs access to fully multi--centered solutions which can be widely separated.}.  

The  point for us is that the required operators in field theory correspond to dynamical fields in the full gravity theory and so exploring them  is a full dynamical problem. In other words, the value of $p$ changes dynamically as a result of the  dynamics of these supergravity fields.  On the other hand it is  the asymptotic values of the fields on the AdS boundary that  have precise  meaning in the field theory, where they are masses and expectation values of field theory operators. The field theory does not know about the full dynamics of the fields in the bulk, although it knows about the value of $p$ through the effective $N$ that sets the number of degrees of freedom.  It is in {\it this} precise sense that $p$, while it has meaning in both theories (connected to pressure in one and number of degrees of freedom in the other), is naturally  dynamical in the gravity theory, while on the field theory side changing  it is rather more akin to motion on the space of field theories, allowing $N$ to change. So $p$ is not dynamical in a particular field theory. This is how the black hole mass $M$ can be energy $U$ in the field theory, and the enthalpy $U+pV$ in the gravity at the same time.

So the stage is set for  how to realize our heat engines. Using the  flow on the space of field theories just described we can perform thermodynamic cycles of the type described in section~\ref{sec:engines},  exploring different values of $p$,  by turning on appropriate choices of operators in the dual field theory. Our heat engines are truly holographic in that we can describe their operation using a dual holographic description in the field theory.

This leads us to  the matter of the mechanical work done over the cycle.  What is the meaning of this work? Normally when we conceive of an engine and the mechanical work it does we have in mind coupling ({\it via} say, a piston) the volume to some external   environment on which we are doing work. This means we must try to understand what the meaning of $V$ is in the field theory. It is not the field theory volume since it has dependence on parameters  other than the AdS scale $\ell$. (In global AdS$_D$, for example, the finite volume the dual field theory is on  is an $S^{D-2}$ of radius $\ell$.) We should look for something analogous to what we saw above with  the pressure, which is a meaningful quantity in the field theory (without actually being a pressure) and was set by some power of  $N$, the  number of degrees of freedom. Our $V$ should be a sort of conjugate to that.  So this implies that it would be a  chemical potential, it seems, for (some positive power of)  the number of degrees of freedom (although since in thermodynamics $p$ is the intensive variable and $V$ the extensive, it may well be that $p$ is more akin to a chemical potential). It would be interesting to identify such a quantity in  field theory. If this is possible,  one's expectation is that  it might  be  a geometrically defined quantity. Important geometrically defined quantities that are related to measures of degrees of freedom in a theory are not unfamiliar in this field. The entanglement entropy is an example\cite{Bombelli:1986rw,Srednicki:1993im}.

Whatever non--volume quantity it is that $V$ turns out to compute in a field theory, it will   get changed  when mechanical work is done by on it one of our heat engines. The picture would be that  field theory {\bf A} is used as a holographic heat engine that can operate on field theory {\bf B} by coupling them together appropriately.  After a cycle, performed by renormalization group  flows in theory {\bf A} as described above, the $V$ (remember, {\it not} volume) of theory {\bf B} has changed. Mechanical work was performed using heat. 

It is probably wise to stop speculating at this point, but it does seem possible that these holographic heat engines may serve as new tools for the study of gauge theories in some enlarged framework yet to be understood.


\section*{Acknowledgements}
 CVJ would like to thank the  US Department of Energy for support under grant DE-FG03-84ER-40168,  the Aspen Center for Physics  for hospitality (under NSF Grant \#1066293) during the preparation of a later version of this manuscript, and Amelia for her support and patience.


\begin{thebibliography}{10}

\bibitem{Bekenstein:1973ur}
J.~D. Bekenstein, ``{Black holes and entropy},''
\href{http://dx.doi.org/10.1103/PhysRevD.7.2333}{{\em Phys.Rev.} {\bf D7}
  (1973)  2333--2346}.

\bibitem{Bekenstein:1974ax}
J.~D. Bekenstein, ``{Generalized second law of thermodynamics in black hole
  physics},''
\href{http://dx.doi.org/10.1103/PhysRevD.9.3292}{{\em Phys.Rev.} {\bf D9}
  (1974)  3292--3300}.

\bibitem{Hawking:1974sw}
S.~Hawking, ``{Particle Creation by Black Holes},''
\href{http://dx.doi.org/10.1007/BF02345020}{{\em Commun.Math.Phys.} {\bf 43}
  (1975)  199--220}.

\bibitem{Hawking:1976de}
S.~Hawking, ``{Black Holes and Thermodynamics},''
\href{http://dx.doi.org/10.1103/PhysRevD.13.191}{{\em Phys.Rev.} {\bf D13}
  (1976)  191--197}.

\bibitem{Caldarelli:1999xj}
M.~M. Caldarelli, G.~Cognola, and D.~Klemm, ``{Thermodynamics of
  Kerr-Newman-AdS black holes and conformal field theories},''
  \href{http://dx.doi.org/10.1088/0264-9381/17/2/310}{{\em Class.Quant.Grav.}
  {\bf 17} (2000)  399--420},
\href{http://arxiv.org/abs/hep-th/9908022}{{\tt arXiv:hep-th/9908022
  [hep-th]}}.

\bibitem{Wang:2006eb}
S.~Wang, S.-Q. Wu, F.~Xie, and L.~Dan, ``{The First laws of thermodynamics of
  the (2+1)-dimensional BTZ black holes and Kerr-de Sitter spacetimes},''
  \href{http://dx.doi.org/10.1088/0256-307X/23/5/009}{{\em Chin.Phys.Lett.}
  {\bf 23} (2006)  1096--1098},
\href{http://arxiv.org/abs/hep-th/0601147}{{\tt arXiv:hep-th/0601147
  [hep-th]}}.

\bibitem{Sekiwa:2006qj}
Y.~Sekiwa, ``{Thermodynamics of de Sitter black holes: Thermal cosmological
  constant},'' \href{http://dx.doi.org/10.1103/PhysRevD.73.084009}{{\em
  Phys.Rev.} {\bf D73} (2006)  084009},
\href{http://arxiv.org/abs/hep-th/0602269}{{\tt arXiv:hep-th/0602269
  [hep-th]}}.

\bibitem{LarranagaRubio:2007ut}
E.~A. Larranaga~Rubio, ``{Stringy Generalization of the First Law of
  Thermodynamics for Rotating BTZ Black Hole with a Cosmological Constant as
  State Parameter},''
\href{http://arxiv.org/abs/0711.0012}{{\tt arXiv:0711.0012 [gr-qc]}}.

\bibitem{Kastor:2009wy}
D.~Kastor, S.~Ray, and J.~Traschen, ``{Enthalpy and the Mechanics of AdS Black
  Holes},'' \href{http://dx.doi.org/10.1088/0264-9381/26/19/195011}{{\em
  Class.Quant.Grav.} {\bf 26} (2009)  195011},
\href{http://arxiv.org/abs/0904.2765}{{\tt arXiv:0904.2765 [hep-th]}}.

\bibitem{Dolan:2010ha}
B.~P. Dolan, ``{The cosmological constant and the black hole equation of
  state},'' \href{http://dx.doi.org/10.1088/0264-9381/28/12/125020}{{\em
  Class.Quant.Grav.} {\bf 28} (2011)  125020},
\href{http://arxiv.org/abs/1008.5023}{{\tt arXiv:1008.5023 [gr-qc]}}.

\bibitem{Cvetic:2010jb}
M.~Cvetic, G.~Gibbons, D.~Kubiznak, and C.~Pope, ``{Black Hole Enthalpy and an
  Entropy Inequality for the Thermodynamic Volume},''
  \href{http://dx.doi.org/10.1103/PhysRevD.84.024037}{{\em Phys.Rev.} {\bf D84}
  (2011)  024037},
\href{http://arxiv.org/abs/1012.2888}{{\tt arXiv:1012.2888 [hep-th]}}.

\bibitem{Dolan:2011jm}
B.~P. Dolan, ``{Compressibility of rotating black holes},''
  \href{http://dx.doi.org/10.1103/PhysRevD.84.127503}{{\em Phys.Rev.} {\bf D84}
  (2011)  127503},
\href{http://arxiv.org/abs/1109.0198}{{\tt arXiv:1109.0198 [gr-qc]}}.

\bibitem{Dolan:2011xt}
B.~P. Dolan, ``{Pressure and volume in the first law of black hole
  thermodynamics},''
  \href{http://dx.doi.org/10.1088/0264-9381/28/23/235017}{{\em
  Class.Quant.Grav.} {\bf 28} (2011)  235017},
\href{http://arxiv.org/abs/1106.6260}{{\tt arXiv:1106.6260 [gr-qc]}}.

\bibitem{Dolan:2012jh}
B.~P. Dolan, ``{Where is the PdV term in the fist law of black hole
  thermodynamics?},''
\href{http://arxiv.org/abs/1209.1272}{{\tt arXiv:1209.1272 [gr-qc]}}.

\bibitem{Altamirano:2014tva}
N.~Altamirano, D.~Kubiznak, R.~B. Mann, and Z.~Sherkatghanad, ``{Thermodynamics
  of rotating black holes and black rings: phase transitions and thermodynamic
  volume},'' \href{http://dx.doi.org/10.3390/galaxies2010089}{{\em Galaxies}
  {\bf 2} (2014)  89--159},
\href{http://arxiv.org/abs/1401.2586}{{\tt arXiv:1401.2586 [hep-th]}}.

\bibitem{Henneaux:1984ji}
M.~Henneaux and C.~Teitelboim, ``{The Cosmological Constant as a Canonical
  Variable},''
\href{http://dx.doi.org/10.1016/0370-2693(84)91493-X}{{\em Phys.Lett.} {\bf
  B143} (1984)  415--420}.

\bibitem{Teitelboim:1985dp}
C.~Teitelboim, ``{The Cosmological Constant as a Thermodynamic Black Hole
  Parameter},''
\href{http://dx.doi.org/10.1016/0370-2693(85)91186-4}{{\em Phys.Lett.} {\bf
  B158} (1985)  293--297}.

\bibitem{Henneaux:1989zc}
M.~Henneaux and C.~Teitelboim, ``{The Cosmological Constant and General
  Covariance},''
\href{http://dx.doi.org/10.1016/0370-2693(89)91251-3}{{\em Phys.Lett.} {\bf
  B222} (1989)  195--199}.

\bibitem{Parikh:2005qs}
M.~K. Parikh, ``{The Volume of black holes},''
  \href{http://dx.doi.org/10.1103/PhysRevD.73.124021}{{\em Phys.Rev.} {\bf D73}
  (2006)  124021},
\href{http://arxiv.org/abs/hep-th/0508108}{{\tt arXiv:hep-th/0508108
  [hep-th]}}.

\bibitem{recentpaper}
P.~C. Tjiang and S.~H. Sutanto, ``{The efficiency of the Carnot cycle with
  arbitrary gas equations of state},''
  \href{http://dx.doi.org/10.1088/0143-0807/27/4/004}{{\em {European Journal of
  Physics}} {\bf 27} (2006)  719--726},
  \href{http://arxiv.org/abs/{physics/0601173}}{{\tt {physics/0601173}}}.

\bibitem{Chamblin:1999tk}
A.~Chamblin, R.~Emparan, C.~V. Johnson, and R.~C. Myers, ``Charged {A}d{S}
  black holes and catastrophic holography,'' {\em Phys. Rev.} {\bf D60} (1999)
  064018,
\href{http://arxiv.org/abs/hep-th/9902170}{{\tt hep-th/9902170}}.

\bibitem{Chamblin:1999hg}
A.~Chamblin, R.~Emparan, C.~V. Johnson, and R.~C. Myers, ``Holography,
  thermodynamics and fluctuations of charged ads black holes,'' {\em Phys.
  Rev.} {\bf D60} (1999)  104026,
\href{http://arxiv.org/abs/hep-th/9904197}{{\tt hep-th/9904197}}.

\bibitem{Kubiznak:2012wp}
D.~Kubiznak and R.~B. Mann, ``{P-V criticality of charged AdS black holes},''
  \href{http://dx.doi.org/10.1007/JHEP07(2012)033}{{\em JHEP} {\bf 1207} (2012)
   033},
\href{http://arxiv.org/abs/1205.0559}{{\tt arXiv:1205.0559 [hep-th]}}.

\bibitem{Maldacena:1997re}
J.~M. Maldacena, ``The large n limit of superconformal field theories and
  supergravity,'' {\em Adv. Theor. Math. Phys.} {\bf 2} (1998)  231--252,
\href{http://arxiv.org/abs/hep-th/9711200}{{\tt hep-th/9711200}}.

\bibitem{Gubser:1998bc}
S.~S. Gubser, I.~R. Klebanov, and A.~M. Polyakov, ``Gauge theory correlators
  from non-critical string theory,'' {\em Phys. Lett.} {\bf B428} (1998)
  105--114,
\href{http://arxiv.org/abs/hep-th/9802109}{{\tt hep-th/9802109}}.

\bibitem{Witten:1998qj}
E.~Witten, ``Anti-de sitter space and holography,'' {\em Adv. Theor. Math.
  Phys.} {\bf 2} (1998)  253--291,
\href{http://arxiv.org/abs/hep-th/9802150}{{\tt hep-th/9802150}}.

\bibitem{Witten:1998zw}
E.~Witten, ``Anti-de sitter space, thermal phase transition, and confinement in
  gauge theories,'' {\em Adv. Theor. Math. Phys.} {\bf 2} (1998)  505--532,
\href{http://arxiv.org/abs/hep-th/9803131}{{\tt hep-th/9803131}}.

\bibitem{Balasubramanian:1999re}
V.~Balasubramanian and P.~Kraus, ``{A stress tensor for anti-de Sitter
  gravity},'' \href{http://dx.doi.org/10.1007/s002200050764}{{\em Commun. Math.
  Phys.} {\bf 208} (1999)  413--428},
\href{http://arxiv.org/abs/hep-th/9902121}{{\tt arXiv:hep-th/9902121}}.

\bibitem{Johnson:2003gi}
C.~V. Johnson, ``{D-branes},'' {\em {Cambridge University Press}} ({2003})  ,
\href{http://arxiv.org/abs/{ISBN-9780521809122}}{{\tt {ISBN-9780521809122}}}.

\bibitem{Dolan:2013dga}
B.~P. Dolan, ``{The compressibility of rotating black holes in
  $D$-dimensions},''
  \href{http://dx.doi.org/10.1088/0264-9381/31/3/035022}{{\em
  Class.Quant.Grav.} {\bf 31} (2014)  035022},
\href{http://arxiv.org/abs/1308.5403}{{\tt arXiv:1308.5403 [gr-qc]}}.

\bibitem{Girardello:1998pd}
L.~Girardello, M.~Petrini, M.~Porrati, and A.~Zaffaroni, ``{Novel local CFT and
  exact results on perturbations of N=4 superYang Mills from AdS dynamics},''
  \href{http://dx.doi.org/10.1088/1126-6708/1998/12/022}{{\em JHEP} {\bf 9812}
  (1998)  022},
\href{http://arxiv.org/abs/hep-th/9810126}{{\tt arXiv:hep-th/9810126
  [hep-th]}}.

\bibitem{Distler:1998gb}
J.~Distler and F.~Zamora, ``{Nonsupersymmetric conformal field theories from
  stable anti-de Sitter spaces},'' {\em Adv.Theor.Math.Phys.} {\bf 2} (1999)
  1405--1439,
\href{http://arxiv.org/abs/hep-th/9810206}{{\tt arXiv:hep-th/9810206
  [hep-th]}}.

\bibitem{Kraus:1998hv}
P.~Kraus, F.~Larsen, and S.~P. Trivedi, ``The coulomb branch of gauge theory
  from rotating branes,'' {\em JHEP} {\bf 03} (1999)  003,
\href{http://arxiv.org/abs/hep-th/9811120}{{\tt hep-th/9811120}}.

\bibitem{Klebanov:1999tb}
I.~R. Klebanov and E.~Witten, ``{AdS/CFT correspondence and symmetry
  breaking},'' \href{http://dx.doi.org/10.1016/S0550-3213(99)00387-9}{{\em
  Nucl. Phys.} {\bf B556} (1999)  89--114},
\href{http://arxiv.org/abs/hep-th/9905104}{{\tt arXiv:hep-th/9905104}}.

\bibitem{Intriligator:1999ai}
K.~A. Intriligator, ``{Maximally supersymmetric RG flows and AdS duality},''
  \href{http://dx.doi.org/10.1016/S0550-3213(99)00803-2}{{\em Nucl.Phys.} {\bf
  B580} (2000)  99--120},
\href{http://arxiv.org/abs/hep-th/9909082}{{\tt arXiv:hep-th/9909082
  [hep-th]}}.

\bibitem{Costa:1999sk}
M.~S. Costa, ``{Absorption by double centered D3-branes and the Coulomb branch
  of N=4 SYM theory},''
  \href{http://dx.doi.org/10.1088/1126-6708/2000/05/041}{{\em JHEP} {\bf 0005}
  (2000)  041},
\href{http://arxiv.org/abs/hep-th/9912073}{{\tt arXiv:hep-th/9912073
  [hep-th]}}.

\bibitem{Costa:2000gk}
M.~S. Costa, ``{A Test of the AdS / CFT duality on the Coulomb branch},''
  \href{http://dx.doi.org/10.1016/S0370-2693(00)00484-6}{{\em Phys.Lett.} {\bf
  B482} (2000)  287--292},
\href{http://arxiv.org/abs/hep-th/0003289}{{\tt arXiv:hep-th/0003289
  [hep-th]}}.

\bibitem{Skenderis:2006uy}
K.~Skenderis and M.~Taylor, ``{Kaluza-Klein holography},''
  \href{http://dx.doi.org/10.1088/1126-6708/2006/05/057}{{\em JHEP} {\bf 0605}
  (2006)  057},
\href{http://arxiv.org/abs/hep-th/0603016}{{\tt arXiv:hep-th/0603016
  [hep-th]}}.

\bibitem{Skenderis:2006di}
K.~Skenderis and M.~Taylor, ``{Holographic Coulomb branch vevs},''
  \href{http://dx.doi.org/10.1088/1126-6708/2006/08/001}{{\em JHEP} {\bf 0608}
  (2006)  001},
\href{http://arxiv.org/abs/hep-th/0604169}{{\tt arXiv:hep-th/0604169
  [hep-th]}}.

\bibitem{Evans:2001zn}
N.~J. Evans, C.~V. Johnson, and M.~Petrini, ``{Clearing the throat: Irrelevant
  operators and finite temperature in large N gauge theory},''
  \href{http://dx.doi.org/10.1088/1126-6708/2002/05/002}{{\em JHEP} {\bf 0205}
  (2002)  002},
\href{http://arxiv.org/abs/hep-th/0112058}{{\tt arXiv:hep-th/0112058
  [hep-th]}}.

\bibitem{Freedman:1999gk}
D.~Freedman, S.~Gubser, K.~Pilch, and N.~Warner, ``{Continuous distributions of
  D3-branes and gauged supergravity},''
  \href{http://dx.doi.org/10.1088/1126-6708/2000/07/038}{{\em JHEP} {\bf 0007}
  (2000)  038},
\href{http://arxiv.org/abs/hep-th/9906194}{{\tt arXiv:hep-th/9906194
  [hep-th]}}.

\bibitem{Brandhuber:1999jr}
A.~Brandhuber and K.~Sfetsos, ``{Wilson loops from multicenter and rotating
  branes, mass gaps and phase structure in gauge theories},'' {\em
  Adv.Theor.Math.Phys.} {\bf 3} (1999)  851--887,
\href{http://arxiv.org/abs/hep-th/9906201}{{\tt arXiv:hep-th/9906201
  [hep-th]}}.

\bibitem{Bombelli:1986rw}
L.~Bombelli, R.~K. Koul, J.~Lee, and R.~D. Sorkin, ``{A Quantum Source of
  Entropy for Black Holes},''
  \href{http://dx.doi.org/10.1103/PhysRevD.34.373}{{\em Phys.Rev.} {\bf D34}
  (1986)  373--383}.

\bibitem{Srednicki:1993im}
M.~Srednicki, ``{Entropy and area},''
  \href{http://dx.doi.org/10.1103/PhysRevLett.71.666}{{\em Phys. Rev. Lett.}
  {\bf 71} (1993)  666--669},
\href{http://arxiv.org/abs/hep-th/9303048}{{\tt arXiv:hep-th/9303048}}.

\end{thebibliography}

\providecommand{\href}[2]{#2}\begingroup\raggedright\endgroup

\end{document}